\documentclass[12pt,preprint]{aastex}

\slugcomment{draft version \today, }

\shorttitle{GRB090529A}
\shortauthors{Xin et al.}

\begin{document}

\title{
{The shallow-decay phase in both optical and x-ray afterglows of \em Swift} GRB 090529A:
Energy injection into a wind-type medium?}

\author{L. P. Xin\altaffilmark{1},
A. Pozanenko\altaffilmark{2},
D. A. Kann\altaffilmark{3},
D. Xu\altaffilmark{4},
J. Gorosabel\altaffilmark{5},
G. Leloudas\altaffilmark{4},
J. Y. Wei\altaffilmark{1},
M. Andreev \altaffilmark{6,7},
S. F. Qin\altaffilmark{1},
M. Ibrahimov\altaffilmark{8},
X. H. Han\altaffilmark{1},
A. de Ugarte Postigo\altaffilmark{4},
Y. L. Qiu\altaffilmark{1},
J. S. Deng\altaffilmark{1},
A. Volnova\altaffilmark{9},
P. Jakobsson\altaffilmark{10},
A. J. Castro-Tirado\altaffilmark{5},
F. Aceituno\altaffilmark{11},
J. P. U. Fynbo\altaffilmark{4},
J. Wang\altaffilmark{1},
R. Sanchez-Ramirez\altaffilmark{5},
V. Kouprianov\altaffilmark{12},
W. K. Zheng\altaffilmark{13},
J. C. Tello\altaffilmark{5},
C. Wu\altaffilmark{1}
}

\altaffiltext{1}{National Astronomical Observatories, Chinese Academy of Sciences, Beijing 100012, China, xlp@nao.cas.cn}
\altaffiltext{2}{Space Research Institute (IKI), 84\/32 Profsoyuznaya str., Moscow, 117997, Russia}
\altaffiltext{3}{Th\"uringer Landessternwarte Tautenburg, Sternwarte 5, 07778, Tautenburg}
\altaffiltext{4}{Dark Cosmology Centre, Niels Bohr Institute, University of Copenhagen, Juliane Maries Vej 30, 2100 Copenhagen, Denmark}
\altaffiltext{5}{Instituto de Astrof\'{\i}sica de Andaluc\'{\i}a (IAA-CSIC), PO Box 3.004, 18.080 Granada, Spain,\#8233}
\altaffiltext{6}{Terskol Branch of Institute of Astronomy of RAS, Kabardino-Balkaria Republic 361605, Russian Federation}
\altaffiltext{7}{International Centre of Astronomical and Medico-Ecological Research of NASU, 27 Akademika Zabolotnoho str., 03680, Kyiv, Ukraine}
\altaffiltext{8}{Ulugh Beg Astronomical Institute, Tashkent 700052, Uzbekistan}
\altaffiltext{9}{Sternberg Astronomical Institute, Moscow State University, Universitetsky pr., 13, Moscow 119992, Russia}
\altaffiltext{10}{Centre for Astrophysics and Cosmology, Science Institute, University of Iceland, Dunhagi 5, IS-107 Reykjavik, Iceland}
\altaffiltext{11}{Observatorio de Sierra Nevada, Instituto de Astrof\'{\i}sica de Andaluc\'{\i}a (IAA-CSIC), PO Box 3.004, 18.080 Granada, Spain,\#8233}
\altaffiltext{12}{Main Astronomical Observatory of RAS, Pulkovo, St.Petersburg, 196140, Russia}
\altaffiltext{13}{Department of Physics, University of Michigan, Ann Arbor, MI 48109, USA}

\begin{abstract}
The energy injection model is usually proposed to interpret the shallow-decay phase in  {\em Swift} GRB X-ray afterglows. However, very few GRBs have  simultaneous signatures of energy injection in their optical and X-ray afterglows. Here, we report optical observations of GRB 090529A from 2000 sec to $\sim$10$^6$ sec after the burst, in which an achromatic decay is seen at both wavelengths. The optical light curve shows a decay from 0.37 to 0.99 with a break at $\sim$10$^5$ sec.
In the same time interval, the decay indices of the X-ray light curve changed
from 0.04 to 1.2.
Comparing these values with the closure
relations, the segment after 3$\times10^{4}$ sec is consistent with the prediction of the forward shock in an ISM medium without any energy injection.
The shallow-decay phase between 2000 to 3$\times10^{4}$ sec could be due to the external shock in a wind-type-like medium with an energy injection
under the condition of $\nu_o < \nu_c < \nu_x$. However, the constraint of the spectral region is not well consistent with the multi-band observations.
For this shallow-decay phase, other models are also possible, such as energy injection with evolving microphysical parameters, or a jet viewed off-axis,etc.
\end{abstract}


\section{Introduction}

Gamma-ray bursts (GRBs) are considered to be produced by the merger of binary compact stars (Type-I GRBs), or by the death of massive stars (Type-II GRBs) (e.g., Paczy\'nski. 1998; Zhang et al. 2009), with a relativistic fireball shell (ejecta) expanding into an uniform interstellar medium (ISM), or the preburst stellar wind of the progenitor star with a density distribution of  $\rho \propto R^{-2}$ (e.g., Chevalier \& Li 1999; Dai \& Lu 1998a). The standard fireball model (Sari et al. 1998) was well capable of interpreting most of observations before {\em Swift}.  After {\em Swift} was launched in 2004 (Gehrels et al. 2004), our understanding of GRB physics has been revolutionized by lots of unexpected discoveries from space- and ground-based observations.  Almost half of the {\em Swift} GRBs show canonical light curves  in their X-ray afterglows,and the origin of shallow-decay behavior is highly debated  (e.g., Granot 2006; Zhang et al. 2006; Liang et al. 2007;  Nousek et al. 2006; Evans et al. 2009). Several models have been proposed to interpret this behavior, such as prior emission (Yamazaki 2009; Liang et al. 2010), off-axis jet viewing (Granot et al. 2002; Eichler \& Granot 2006),  X-ray dust scattering (Shao \& Dai 2007), microphysical parameters evolving (Granot, Konigl \& Piran 2006; Panaitescu et al. 2006) and energy injection (Dai \& Lu 1998b,1998c; Sari \& M{\'e}sz{\'a}ros 2000;  Nousek et al. 2006; Zhang et al. 2006;).

The signature of energy injection has been reported in the
literature for a few GRB afterglows.
 For example,
GRB 050408 (de Ugarte Postigo et al. 2007),
whose multi-band light curve shows a rebrightenning phase at 2.9 days after the burst accromatically,
is considered as a likely off-axis event with an energy injection.
This case is also thought to have taken place in both X-ray and optical afterglows  of GRB 071010A (Covino et al. 2008),
whose sharp rebrightenning at 0.6 days after the burst  is considered as a discrete energy injection.
The plateau phases in both the X-ray and the optical afterglows of GRB 061121 are consistent
with an injection of energy (Page et al. 2007),
but the transition from plateau to the later phase is only visible in X-ray afterglows.
A long-duration shallow-decay phases in both the X-ray and the optical afterglows is also
evident in long-duration GRB 060729 (T$_{90}= 105$ sec) (Grupe et al. 2007),
which could be well explained via a long-term energy injection.
The transition break between the shallow-decay phase and the normal-decay phase lies at about 60 ks and about 50 ks for X-ray and optical afterglows,
respectively. There are multiple further examples of GRBs which feature strong energy injections in the
optical, but which are hardly or not at all visible in the X-rays (usually due to sparse
coverage), such as GRB 021004 (e.g., de Ugarte Postigo et al. 2005), GRB 030329 (e.g.,
Lipkin et al. 2004), GRB 060206 (Wozniak et al. 2006, Monfardini et al. 2006), GRB 060526
(Dai et al. 2007, Th\"one et al. 2010), GRB 070125 (Updike et al. 2008) and GRB 090926A
(Rau et al. 2010, Cenko et al. 2011).

GRB 090529A provides another good case, like GRB 060729, that both optical and X-ray afterglows have a long-term  energy injection
before the phase of normal achromatic decay.
 Thus, it is a good example for us to undertake a detailed study of this topic. Both X-ray and optical afterglows show shallow-decay segments within a similar time range, which is followed by a steeper, almost achromatic decay. This behavior can be well-interpreted as an energy
injection via a refreshed shock or late-time central engine activity.

In this paper, we report our observations of the optical afterglow for GRB 090529A using several ground-based optical telescopes. The optical and X-ray afterglow data are used to explore the nature of this event. Our observations are reported in section 2. A joint optical and X-ray data analysis and discussions are presented in section 3. Energy budget will be given in section 4.
Summary and conclusions are presented in section 5. The notation $f_\nu\propto t^{-\alpha}\nu^{-\beta}$ is used throughout the paper, where $f_\nu$ is the spectral flux density at the frequency $\nu$.

\section{Observations and Data Reduction}

GRB 090529A was detected by the Burst Alert Telescope (BAT) onboard of {\em Swift} (Sakamoto et al. 2009a) at 14:12:35 UT on 2009 May 29 (trigger 353540)
 during a pre-planed slew. Since {\em Swift} did not observe the start of the burst,
its $T_{90}$ duration could only be given as a lower limit, but it
was still longer than 100s in the 15-350 KeV band (Markwardt et al. 2009),
and it therefore belongs to a long-duration class of bursts (Kouveliotou et al. 1993).
It is interesting that no other missions detected this burst. Even SPI-ACS/INTEGRAL which could detect  the burst have no rate increase in the data around GRB 090529 trigger time. Most probably the burst was too soft to be detected by SPI-ACS. The significant short spike with a duration of 0.1 s occured 20 s before trigger time\footnote{http://www.isdc.unige.ch/integral/science/grb}, however£¬ there is no clear relationship of the spike with GRB 090529.
The {\em Swift} X-ray Telescope (XRT) began to observe the burst 197.1 seconds after the burst, and found a counterpart. Meanwhile, no new source was found in the first white finding chart of the {\em Swift} UVOT instrument, but a new source was detected in the second white image at 883 sec after the burst trigger (Sakamoto et al. 2009a). The optical afterglow was observed by several ground-based telescopes, and its spectroscopic redshift of $z=2.63$ was determined by the ESO Very Large Telescope (Malesani et al. 2009). The time-integrated $\gamma$-ray spectrum is well fitted by a single power-law with a photon index of 2.0$\pm$0.3 (Markwardt et al. 2009). The fluence in the 15-150 keV band is 6.8$\pm$1.7$\times$10$^{-7}$ erg cm$^{-2}$.
The isotropic energy E$_{iso}$ would be estimated to be about $7\times10^{52}$ erg.
With the relation between photon index $\Gamma$ and observational peak energy
E$_{peak,o}$ obtained by simulations (Sakamoto et al. 2009),
peak energy E$_{peak,o}$ in the observational frame  could be inferred as $40\pm23$ KeV.
Therefore, this burst would be consistent with the Amati relationship (Amati et al. 2002) as most of
long duration GRBs .

\subsection{Swift/XRT X-ray Afterglow }

The {\em Swift}/XRT light curve and spectrum are extracted from the UK Swift
Science Data Centre at the University of Leicester (Evans et al.
2009)\footnote{http://www.swift.ac.uk/results.shtml}.
We also fit the X-ray spectrum
with the $Xspec$ package.
The time-integrated X-ray spectrum from
all the Photon Counting (PC) mode data after
 3 ks-3$\times10^{4}$ s
is well fitted by an absorbed
power-law model, with a photon power-law index $\Gamma=1.6^{+0.54}_{-0.37}$.
No significant host $N_H$ excess over the Galactic value
is detected.
The time-integrated X-ray spectrum after 3$\times10^{4}$ s post the trigger can be fit with
a power-law model with a photon power-law index of $\Gamma=2.53^{+1.45}_{-0.96}$.
The two $\Gamma$ values agree with each other within error due to large uncertainties,
so there is no significant evidence for spectral evolution.

\subsection{Optical Afterglow}

The 0.8-m Tsinghua University - National Astronomical Observatories
Telescope (TNT) is located at Xinglong observatory in China.
GRB 090529A  was observed
by TNT  starting from  May 29, 14:36:54 (UT),
24.3 min after the burst. A series of $clear$ (C$_R$) and $R$ band images were obtained.
This led to the identification of the optical counterpart.

The Maidanak AZT-22 (1.5m) telescope is located at Maidanak Observatory, in south-east of the Republic of Uzbekistan.
It observed the optical afterglow of GRB 090529A from 0.15 days to 0.2 days for several times, and
a series of images in the $R$ band were obtained. The optical counterpart was detected in all the single image.

The 1.34m Schmidt telescope of the
Th\"uringer Landessternwarte Tautenburg (TLS) in Germany,
began to observe the optical afterglow at
6.6 hours after the  burst.
Six dithered  $I_C$ band images with an exposure time of 300 s for each single image, and six images in the $R_C$ band with an exposure
time of 600 s each, were obtained.

The optical afterglow was observed by the Zeiss-600 telescope (Zeiss-600) of Mt.Terskol
observatory in $R$-band filter between May 29 19:40 -- 20:20 (UT).
The optical counterpart of this burst could
be detected in a combined images with a total time exposure of 2400 sec.

The optical afterglow was also observed by MTM-500 telescope of
Kislovodsk solar station of the Pulkovo observatory in $R$-band with several
series on May 29 between 19:58 -- 22:20 (UT). The optical counterpart was
detected in combined images.

The 1.5-m OSN
telescope carried out $R$-band observations of the GRB 090529A
on May 29.87-29.93 UT, 0.27-0.33 days after the GRB.
The optical afterglow was detected in the combined image with a total exposure time of 4800 sec.

The Nordic Optical Telescope (NOT), carried out a multiple follow-up
observations of GRB 090529A starting 1.29 days after the burst.
the multiple follow-up observations up to 6 days post the trigger.
Several $R$-band images were obtained. Each of these images has an
exposure time of 600 sec.
The optical afterglow was well detected in these images.

Data reduction was carried out following standard routines in the
IRAF\footnote{IRAF is distributed by NOAO, which is operated by AURA, Inc., under
cooperative agreement with NSF.} package, including bias and flat-field
corrections. Point spread function (PSF)
photometry was applied via the DAOPHOT task in the IRAF package to obtain the
instrumental magnitudes. During the reduction, some frames were combined in
order to increase the signal-to-noise ratio (S/N). In the calibration and analysis,
the TNT $C\_{R}$ band was treated as the $R$ band, because they are similar with
each other within uncertainties of 0.07 mag (Xin et al. 2010). TLS images were
reduced in a standard fashion and analyzed under
MIDAS\footnote{http://www.eso.org/sci/software/esomidas/} using seeing-matched
aperture photometry.
Absolute calibration was performed using the Sloan Digital Sky Survey (SDSS,
Adelman-McCarthy et al. 2008), with a conversion of SDSS to the Johnson-Cousins
system (Lupton 2005)
{\footnote{http://www.sdss.org/dr6/algorithms/sdssUBVRITransform.html\#Lupton2005}.
The optical data of GRB 090529A observed by TNT,Maidanak, TLS, Z-600, OSN,
and NOT are reported in Table ~\ref{obslog}.
For completeness, the R-band data from GCN Circular 9487 (Balman et al. 2009) is  re-calibrated  with  SDSS reference stars
and presented here.

\begin{deluxetable}{lccccc}
\tabletypesize{\scriptsize}
\tablecaption{Optical afterglow photometry log of GRB 090529A. The
reference time $T_0$ is {\em Swift} BAT burst trigger time of 14:12:35 UT.
The data have not been corrected for the Galactic extinction ($E_{B-V}=0.021$, Schlegel et al.1998) \label{obslog}.
The first column of this table is the mean time. }
\tablewidth{0pt}
\startdata
\hline\hline
T-T$_{0}$(min)   &  Exposure(sec)    &  Filter  &  Magnitude     &  Mag$\_$Err   &  Telescope  \\
\hline
25.825    &  160   &  C$_R$   &  19.84  &  0.14   &  TNT          \\
29.627    &  240   &  C$_R$   &  19.92  &  0.12   &  TNT          \\
37.230    &  600   &  R   &  20.05  &  0.12   &  TNT          \\
47.843    &  600   &  R   &  19.98  &  0.11   &  TNT          \\
58.427    &  600   &  R   &  19.99  &  0.11   &  TNT          \\
69.025    &  600   &  R   &  20.18  &  0.11   &  TNT          \\
79.610    &  600   &  R   &  20.47  &  0.13   &  TNT          \\
100.368   &  1800  &  R   &  20.22  &  0.10   &  TNT          \\
131.198   &  1800  &  R   &  20.43  &  0.11   &  TNT          \\
162.030   &  1800  &  R   &  20.64  &  0.14   &  TNT          \\
218.548   &  4800  &  R   &  20.66  &  0.15   &  TNT          \\
223.330   &  300   &  R   &  20.73  &  0.07   &  Maidanak     \\
229.795   &  300   &  R   &  20.70  &  0.05   &  Maidanak     \\
235.987   &  300   &  R   &  20.65  &  0.05   &  Maidanak     \\
242.597   &  300   &  R   &  20.75  &  0.06   &  Maidanak     \\
248.083   &  300   &  R   &  20.59  &  0.05   &  Maidanak     \\
253.583   &  300   &  R   &  20.80  &  0.06   &  Maidanak     \\
259.070   &  300   &  R   &  20.72  &  0.07   &  Maidanak     \\
265.420   &  300   &  R   &  20.76  &  0.06   &  Maidanak     \\
270.907   &  300   &  R   &  20.75  &  0.07   &  Maidanak     \\
276.380   &  300   &  R   &  20.80  &  0.06   &  Maidanak     \\
281.880   &  300   &  R   &  20.72  &  0.06   &  Maidanak     \\
287.367   &  300   &  R   &  20.76  &  0.06   &  Maidanak     \\
294.020   &  300   &  R   &  20.73  &  0.06   &  Maidanak     \\
299.505   &  300   &  R   &  20.76  &  0.06   &  Maidanak     \\
347.415   &  2400  &  R   &  20.94  &  0.23   &  Zeiss-600    \\
405.648   &  3600  &   R  &  20.70  & 0.25  &   MTM-500$^a$ \\
478.656   &  3600  &   R  &  21.30  & 0.40   &  MTM-500$^a$ \\
414.535   &  1800  &  Ic  &  20.27  &  0.09   &  TLS          \\
432.000   &  4800  &  R   &  21.01  &  0.05   &  OSN          \\
438.722   &  600   &  Rc  &  21.11  &  0.13   &  TLS          \\
449.555   &  600   &  Rc  &  21.13  &  0.13   &  TLS          \\
460.372   &  600   &  Rc  &  21.23  &  0.43   &  TLS          \\
596.160   &  900   &  R   &  20.90  &  0.20   &  GCN9485$^b$  \\
663.462   &  1800  &  Rc  &  21.00  &  0.11   &  TLS          \\
1874.800  &  600   &  R   &  21.81  &  0.05   &  NOT          \\
1885.750  &  600   &  R   &  21.77  &  0.05   &  NOT          \\
1896.550  &  600   &  R   &  21.88  &  0.05   &  NOT          \\
4774.000  &  1800  &  R   &  22.79  &  0.12   &  NOT          \\
9145.420  &  1800  &  R   &  23.06  &  0.20   &  NOT          \\
\enddata
\tablecomments{
$a$: The photometry supersedes result reported in GCN circ. 9611 (Volnova et al.
2009).
$b$: This data is derived from the literature (Balman et al. 2009).
We re-calibrated its brightness with  SDSS reference stars.}
\end{deluxetable}

\section{Analysis and discussion}

\subsection{Light curves}

We correct the extinction of our Galaxy ($E_{B-V}=0.023$, Schlegel et al.1998),
and then plot the  optical R-band light curve  in Fig.~\ref{OptLC}.
First, we fit the R-band light curve  after 1000 sec post the burst with a simple single power law $f\sim t^{-\alpha_{pl}}$. The
temporal slope $\alpha_{pl}$ would be $0.5\pm0.01$. The reduced $\chi^2$ is 2.47 with 32 degrees of freedom.
We then fit the same data with a smoothly broken
power-law model  yielding two decaying slopes of $0.37\pm0.03$ and $0.99\pm0.12$ respectively, with a broken time of $\sim9.5\times10^{4}$ s. The reduced $\chi^2$ is 1.08 with 30 degrees of freedom, as shown in Eq. ~\ref{Equ.fit}.
Thus, the smooth broken power-law model is better than the single power-law model to fit the optical light curve.
\begin{equation}
\label{Equ.fit}
F=F_0\left [
\left (   \frac{t}{t_b}\right)^{\omega\alpha_1}+\left (
\frac{t}{t_b}\right)^{\omega\alpha_2}\right]^{-1/\omega}
\end{equation}

\begin{table}
\caption{X-ray
and optical light curves after 3000 sec post burst are fitted with a smoothly broken power-law model.
$^*$In units of $\times$10$^{-13}$ erg cm$^{-2}$ s$^{-1}$.}
  \label{Tab:fit}
  \begin{center}
  \begin{tabular}{ccccccc}
  \hline\noalign{\smallskip}
Interval  & $F_0$ &  $t_b$(s)   & $\alpha_1$ & $\alpha_2$ & $\omega$(fixed) &$\chi^2/dof$\\
   \hline\noalign{\smallskip}

X-Ray  & 7.396$\pm$2.371$^{*}$ & 40095$\pm$17760 & 0.04$\pm$0.20 & 1.17$\pm$0.17 & 3  & 61/59 \\

Optical  &  8.10$\pm$1.27 & 95042$\pm$23457 & 0.37$\pm$0.03 &  0.99$\pm$0.12 & 3 & 32.48/30\\

  \noalign{\smallskip}\hline
  \end{tabular}
\end{center}
\end{table}

{\em Swift} UVOT-white band data  (Sakamoto et al. 2009b) are also plotted in Fig.~\ref{OptLC}.
In order to compare R-band data with {\em Swift} UVOT data, we shift the UVOT white-band data to the
flux level of our Johnson R-band data after 1000 sec post burst.
We find that the R-band and UVOT light curves trace each other well after 1000 sec.
Assuming they also trace each other well before 1000 sec,
we could infer the behavior of the R-band light curve before 1000 sec
from the properties of the UVOT white-band light curve.

The UVOT data at the early phase shows an upper limit at the first observation,
and a positive detection at the second data point.
The latter one is apparently brighter than the previous one.
Other later data  became fainter than the second data for about 0.5 mag, and decayed continuously
with a shallow decay slope of $\sim$0.3.
Just considering the second and the third data of UVOT observations,
the decay index between these two measurements is about 1.24$\pm0.3$, which is not likely to be the emission
from reverse shock, like GRB 990123 (Akerlof et al. 1999).
The existence of the brightest flux of the second white data point,
implies that  the onset of the early afterglow or an early optical flare
have taken place before the shallow-decay phase.
However, owing to the sparse data, the origin of the early UVOT white observations is not clear.

{\em Swift} X-ray light curve is also plotted and fitted in the Fig.~\ref{OptLC}. The X-ray afterglow shows
a {\em Swift} canonical X-ray light curve with
a steep decay during the first observations up to about 3000 sec, and then turning into the shallow-decay phase
with a decay index of 0.04$\pm0.20$, which is followed by a normal decay with a decay slope of 1.17$\pm0.17$.
All these fit parameters are summarized in Table ~\ref{Tab:fit}.

We note that GRB 090529A was triggered in image mode, and was already ongoing when it
cames into the field of view during a pre-planned slew. This means that there was emission
prior to the trigger time for at least 50 s. Therefore, $T_0$ might be shifted to an earlier
time by at least 50 s relative to the trigger time, which consequently affects the decay
slopes of the afterglow emission. However, we have checked that even we shift $T_0$ to 50
s before the trigger time, our fitting results are not affected a lot, since the start time
of the shallow-decay phase is much later, about $\sim10^3$ s after the burst.

\begin{figure}
\epsscale{1}
\plotone{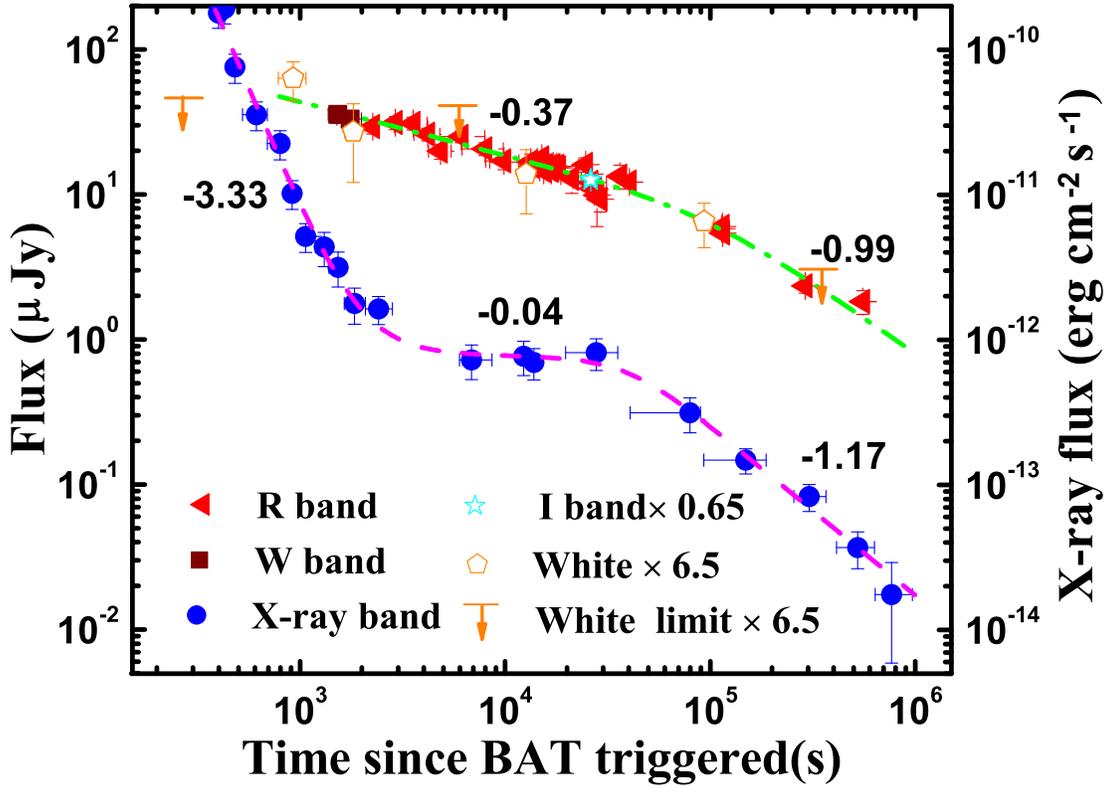}
\caption{Optical and X-ray light curves of GRB 090529A. The optical light curve can be fit with a smoothly broken
power law model with an index transition from $\alpha_{O1}=0.37\pm0.03$ to $\alpha_{O2}=0.99\pm0.12$. The X-rays show a canonical
light curve with a initial steep decay phase, and then a shallow-decay phase with an index of $\alpha_{X1}=0.04\pm0.20$,
which is followed by a normal decay phase with an index of $\alpha_{X2}=1.17\pm0.17$.
{\em Swift} UVOT $white$-band data are also plotted and shifted in flux density for  completeness and comparison.
\label{OptLC}}
\end{figure}

\begin{figure}
\epsscale{1}
\plotone{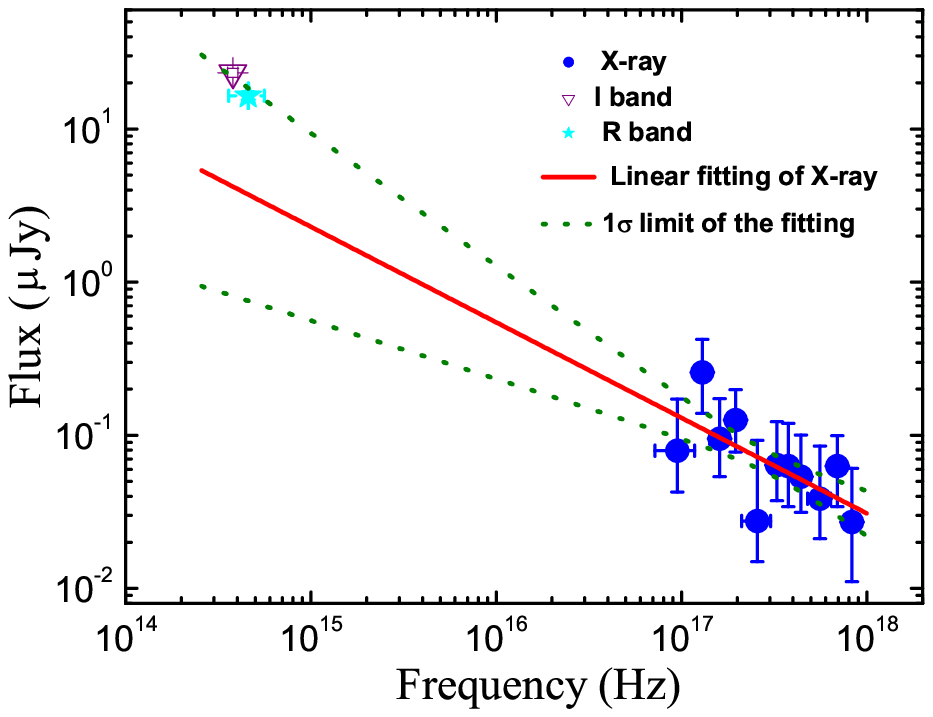}
\caption{The spectrum of X-ray afterglow for GRB 090529A in the time range of 3ks - 30ks sec after the burst.
The optical data points ($R-$ and $I-$band) are labeled with a five-pointed star and a triangle point, respectively.
\label{spec}}
\end{figure}

\subsection{Normal-decay phase}

In the normal decay phase of the X-ray afterglow, the emission process  is usually
in the slow-cooling regime.
For the case of GRB 090529A,
the temporal decay index $\alpha_x$ and spectral index $\beta_x$
are $1.17\pm0.17$ and $1.53^{+1.45}_{-0.96}$, respectively.
These values are roughly consistent with the predictions of the external forward shock model
for the spectral regime $\nu_x >$ max$(\nu_c,\nu_m)$ (Liang et al. 2008).
Additionally,
the X-ray light curve is steeper  than the optical one by $\delta\alpha\sim0.2$. It indicates that the external medium is ISM and
 $\nu_o <\nu_c <\nu_x$ during the power-law decay phase.
In this regime, the electron energy distribution index $p$ could be estimated with the relation of $p=(4\alpha_x+2)/3 \sim 2.2$.
It is consistent with the typical value ($p\sim2.36$) for the electron energy distribution index in GRB model (Curran et al. 2010).
However, since the larger uncertainties of decay slopes, the difference between X-ray and optical afterglows of $\delta\alpha\sim0$
could not be excluded.



\subsection{Shallow-decay phase}

\subsubsection{Spectral index}

In order to investigate the origin of the shallow-decay segments in both optical and X-ray afterglow,
the spectral index from optical to X-ray would be helpful.
To obtain the intrinsic optical flux, the extinction caused by our
Galaxy in the direction of the bursts corrected for,
which is
$E_{B-V}=0.021$ (Schlegel et al.1998), corresponding to $A_{R}=0.055$ mag and $A_{I}=0.040$.
The extinction from the GRB host is not considered here.
The extinction-corrected magnitudes are converted into flux densities.

For the $R$-band data, we inter-plot it with the best fitting of those data during this phase.
For the $I$-band data, we extra-plot it with the fitting result of $R$-band light curve assuming that
the decay indices of light curves in both frequencies are the same in the shallow phase.
The time we calculated for the optical bands is about 15 ks after the burst.
The X-ray spectrum during the shallow-decay phase with its
power-law index and its errors as well as the R-band flux density are plotted in Fig. 2. It is
evident that the optical flux is located in the extrapolation of the X-ray spectral index within
$1\sigma$ errors.
The optical (R-band) to X-ray spectral index $\beta_{OX}$ is very similar to the observed X-ray spectral index of $\sim0.6$.
This fact suggests that the extinction from host galaxy could be negligible.

\subsubsection{Classical external-shock model?}

According to the classical external-shock model (Meszaros, Rees \& Wijers 1998; Sari, Piran \& Narayan 1998;
Sari, Piran \& Halpern 1999; Chevalier \& Li 2000; Dai \& Cheng 2001; Zhang et al. 2006),
Optical and X-ray afterglows have closure relations under several conditions.
We notice that the optical light curve decays steeper than the X-ray one by about
$\delta\alpha \sim 0.37$ at the shallow-decay phase, this
is consistent with the condition of wind-like type medium
in the spectral region of  $\nu_o<\nu_c<\nu_x$ .
In detail,
assuming that the medium has a density distribution of  $\rho \propto R^{-s}$,
the decay indices of optical and X-ray emission would have a relation under the condition of no energy injection,
$\alpha_{o}-\alpha_{x}=-0.25+s/(8-2s)$ (Urata et al. 2007) in the classical external-shock model
for the case of  $\nu_o<\nu_c<\nu_x$ .
With these parameters
($\alpha_o \sim 0.37\pm0.03$ and $\alpha_x \sim 0.04\pm0.20$) of GRB 090529A in this phase,
we can infer that the density parameters {\em s} $\sim$ 2.  This is consistent with the theoretical
prediction of a wind-type medium ($\rho\propto R^{-2}$).
Therefore, it seems that the environment of shallow-decay phase in both X-ray and optical emissions
should be a wind-type medium.
However, the requirement of spectral indices between optical and X-ray emission seems not be  consistent with the result of Fig.2.
A possible solution for this conflict is that the cooling frequency should be marginally above the optical band.
But consequently, the optical spectral index  would be about $\beta_{opt}=0.1^{+0.54}_{-0.37}$
according to the relation of $\beta_{opt} = \beta_X - 0.5$.
However, it seems in disagreement with the $I_C$ band data point from TLS observation.

\subsubsection{Energy injection?}

Considering the shallow-decay indices, continuous energy injection (Dai \& Lu 1998b;
Sari \& M{\'e}sz{\'a}ros 2000;  Nousek et al. 2006; Zhang et al. 2006)  might be occuring,
in which new energy is injected into the forward shock and forms
"refreshed shocks", which makes the temporal decay indices shallower.
This can be produced either  by the long-lasting
activity of the central engine, with the transformation of the Poynting flux energy into kinetic
energy in the external shock (Dai \& Lu 1998b), or by
a brief and short-duration central engine activity, which ejects shells with a
range of Lorentz factors such that slower ejecta catches up
with the decelerated shock (Zhang et al. 2006 and reference therein).

For the first case of Poynting flux energy injection,
the injected luminosity is given by $L(t) = L_0 (t/t_b)^{-q}$ (Zhang et al. 2006).
Given the closure relationship for $\alpha$ and $\beta$ of afterglow in various GRB models (Zhang et al. 2006),
and for the measured values of the X-ray afterglow of GRB 090529A,
$\alpha_x=0.04\pm0.02$, and $\beta_x=0.6^{+0.54}_{-0.42}$,
we estimate that the energy injection index is $q=2(\alpha+1-\beta)/(2+\beta) \sim 0.34\pm0.45$
for $\nu_x > \nu_c$ in a wind medium and the slowing cooling phase.
Because the optical frequency should be located at the spectral region $\nu_{opt}<\nu_c<\nu_x$, the closure relation
via the energy injection would be  $q=2(\alpha-\beta)/(1+\beta)$ for the optical band. In this case,
the energy-injection parameter {\em q} would be $0.5\pm0.5$ for the optical indices
of $\alpha_{opt}=0.37\pm0.03$ and $\beta_{opt} = 0.1^{+0.54}_{-0.37}$.
The two values of {\em q} for both optical and X-ray afterglows are consistent with each other within their uncertainties, implying that
the new-formed energy is injected into the forward shock, making the light curve shallower for both X-ray and optical bands
simultaneously.

For the second case of a range of the Lorentz factors in the  ejecta which follow the form of
$M (> \gamma) \sim \gamma^{-s}$ ($ s >1$) (Zhang et al. 2006),
with our measured value of $q\sim0.4$ (the average value for the optical and X-ray bands),
the Lorentz factor index {\em s} would be about 7 for  the wind model ( Zhang et al. 2006; Nousek et al. 2006).
This result corresponds to the total energy in the fireball  decreasing
as $E_{iso}\propto \gamma^{1-s} \sim \gamma^{-6}$.

It is likely that the simultaneous shallow-decay phase of GRB 090529A in both the X-ray and
optical afterglows could be interpreted by energy injection in a wind-type medium via by a long-lasting central-engine
activity or a range of Lorentz factors in the eject. Meanwhile, the normal decay phase
could be well-explained by the standard forward shock in an ISM medium. Therefore, It seems that
the break from the shallow to the normal decay phase may be caused by the cessation of energy injection, and
simultaneously, the external medium would also have a transition from a wind-type to constant-density ISM.
However it might be unlikely that these two different processes take place at the same time.

\subsubsection{Other models}

{\em Microphysical parameters evolving.}
Panaitescu et al. (2006) generalized the formulae of the synchrotron-shock model by including the variations of some
physical parameters in the
blast wave: the energy injection $E(>\Gamma) \propto \Gamma^{-e}$, the energy ratio for the magnetic field $\epsilon_{B} \propto \Gamma^{-b}$,
the energy ratio for electrons  $\epsilon_{i} \propto \Gamma^{-i}$, and the ambient medium density $n(r) \propto r^{-s}$.
As shown in equations (9) and (10) in Panaitescu et al. (2006), the decay indices for optical and X-ray light curves in the spectral region
$\nu_o<\nu_c<\nu_x$ are derived.
A general relationship between $\alpha_{o}$ and $\alpha_{x}$ can be expressed  by (Urata et al. 2007):
\begin{equation}
 \alpha_o - \alpha_x = \frac{s}{8-2s} - \frac{1}{4}+\frac{3-s}{e+8-2s} \left[ \left( \frac{s}{8-2s}-\frac{1}{4} \right)e - \frac{3}{4}b \right]
\label{aoaxsy}
\end{equation}
which is independent on {\em p} and {\em i}, but has some dependence on {\em e} and {\em b}.
Assuming $s=0$ (ISM case), for the case of GRB 090529A, we would get $(e+3b)/(e+8)=-0.72\times4/3 <0$, which requires that at least one of {\em e} or {\em b} would be negative, which is an unphysical condition.
However,
if we assume that $s=2$ (wind-type medium), we would get $0.72e-3b=1.76$. It means that
when new energy is injected into the wind-type medium, the microphysical parameters
are not only evolving,
but also the energy injection faction and the energy ratio for the magnetic field should have a relationship ( $0.72e-3b=1.76$ )
to meet the case of GRB 090529A.

{\em Off-axis viewing jet model}.  Another  explanation for the plateau phase is that the
afterglow is being observed from viewing angles slightly outside the jet.
Depending on the jet structures, dynamics, viewing angle (e.g. Eichler \& Granot 2006; Marshall et al. 2011),
the light curve can have a long shallow-decay phase at early times or an initial
rising phase after the prompt emission. It is generally consistent with the case of GRB 090529A at the early phase,
except for the earliest detections of a flare-like or
rebrighting-like feature by UVOT. However, the earliest feature might be produced by an additional
physical ingredient, like an abrupt energy injection, before the forward shock
emission from off-axis jet. Thus, this model is also possible.

{\em Dust scattering model}. The dust scattering model was proposed to explain the shallow-decay phase
of {\em Swift} GRBs (Shao \& Dai 2007), like in the optically dark GRB 090417B (Holland et al. 2010).
However, one prediction of this model is that the spectral index would be softer by about
$\Delta\beta = 2\sim3$ in the dust-scattering procedure (Shen et al. 2009) across the break from shallow-decay segment
to the normal-decay phase.
This prediction is not consistent with the case of GRB 090529A. Thus,
this model could be excluded.

{\em Photosphere $+$ external shock model.}  Wu \& Zhang (2011) recently proposed that the X-ray afterglow is dominated by the photospheric emission of a long-lasting wind, while the optical afterglow is dominated by synchrotron emission of the forward shock. Within this model, if the long-lasting central engine has a break in the luminosity time history (e.g. Dai \& Lu 1998b; Zhang \& M\'esz\'aros 2001), both X-ray and optical lightcurves can show an achromatic break at the time of the central engine break, if the total energy injected from the long-lasting engine exceeds that in the blastwave at early times. This model could interpret the data of GRB 090529A,
but it requires that optical and X-ray emissions of this burst should be  dominated by totally different processes.

\section{Energy budget}

Gamma-ray bursts are believed to be produced by ultra-relativistic outflows which are collimated into narrow jets ( see e.g. the review of
Granot \& Ramirez-Ruiz 2011). A break in the multi-wavelengthes light curves is expected when the Lorentz factor $\gamma$ drops below
the inverse of the angular width of the jet $\theta$ (Rhoads 1999; Sari,Piran \& Halpern 1999).
For the case of GRB 090529A, this kind of jet break is not observed as shown in Fig.~\ref{OptLC}, indicating that
the time of a possible jet break should be later than
the last observation of X-ray afterglow, $t_{jet,s}>10^{6}$ sec.
Following Sari et al. (1999), the jet half-opening angle $\theta$ for an ISM environment could be estimated:
$\theta = 0.161\times (t_{jet,d}/(1+z))^{3/8} \times
(n\times\eta_{\gamma}/E_{\gamma,iso,52})^{1/8}$,
where z is the redshift, $t_{jet,d}$ is the break time in days.
The $\theta$ is not sensitive to the isotropic energy E$_{iso}$.
Under an assumption of the
typical values: $n=0.1$ cm$^{-3}$ and $\eta_{\gamma}=0.2$ (Gao \& Dai 2010),
the opening angle of jet $\theta$ should be larger than 6.84 degree.
This half-opening angle is larger than those for most of GRBs (Gao \& Dai 2010).
With the relation of
$E_{\gamma}=(1-cos\theta)E_{\gamma,iso}$,
$E_{\gamma}$ is estimated to be about 5$\times10^{50}$ ergs.

\section{Summary and conclusion}
We report the optical observations of GRB 090529A by various ground-based telescopes.
The optical light curve shows a shallow-decay phase from $\sim 2000$ sec after the burst with a decay index of about 0.37,
which is followed by a "normal" decay segment starting at about $10^{5}$ sec after the  burst.
Before the shallow-decay phase in the optical afterglow,  a tentative flare-like event or rebrightening exists
at earlier times  ($<$2000s) according to the
observations of the UVOT instrument.
Meanwhile, the X-ray afterglow shows a canonical light curve. The shallow-decay phase in the X-ray afterglow starts
at about 3000 sec after the burst, and ends at $\sim 10^{5} $ sec, showing a decay index of about 0.04.
The later phase after the break ($>10^{5}$ s) in both X-ray and optical afterglows could be
explained well with the standard forward-shock model without any energy injection
in the case of $\nu_x >$ max$(\nu_c,\nu_m)$ in an ISM medium.
During our observations, no any jet break signature is detected. The time for jet break
should be later than the end time of X-ray observations ($\sim10^{6}$s),
indicating that the half-opening angle of jet is larger than 6.84 degree,
and the collimated-corrected energy is about 5$\times10^{50}$erg,
making this burst being consistent with the Ghirlanda-relation .

To model the simultaneous shallow-decay phase in both the X-ray and the optical afterglows,
energy injection into a wind-type medium are needed, but there is also a constraint of
the spectral condition of $\nu_m < \nu_o < \nu_x < \nu_c $,
which is not well consistent with the observations of optical multi-band observations.
Besides, the transition break between the shallow-decay phase and the normal-decay phase should be not only due to
the cessation of energy injection, but also correspond to the transition of the external medium density
from a wind-type medium to an ISM.
However it might be unlikely that these two different processes take place at the same time.
Other models, like  microphysical parameters evolving and off-axis viewing jet model, are also possible for the shallow-decay phase.

\acknowledgments
We thank the anonymous referee  for comments, which helped us to improve the paper.
We also wish to thank Bing Zhang for the useful discussions on the modelings of this burst.
This work made use of data supplied by the UK Swift Science Data Centre
at the University of Leicester.
This work is supported by the Young Researcher Grant of National Astronomical Observatories,
Chinese Academy of Sciences.
Xin acknowledges the support by NSFC  grant  11103036.
 Wu acknowledges the support by NSFC grant 10903010.
DAK thanks U. Laux for help with obtaining the observations and B. Stecklum for observing
time. AP and AV acknowledge support by RFFI grants 10-07-00342\_a and 11-01-92202Mong\_a.

\end{document}